\documentstyle[12pt]{article}
\newcommand{\be}{\begin{equation}}
\newcommand{\vom}{\omega^{\scriptstyle *}}
\newcommand{\ee}{\end{equation}}
\newcommand{\bea}{\begin{eqnarray}}
\newcommand{\eea}{\end{eqnarray}}
\newcommand{\bm}[1]{\mbox{\boldmath${#1}$\unboldmath}}
\newcommand{\td}[1]{{\large\tilde{#1}}}
\newcommand{\bel}[1]{\be\label{#1}}
\newcommand{\re}[1]{Eq.~(\ref{#1})}
\newcommand{\mbs}[1]{\mbox{$\scriptstyle{#1}$}}
\newcommand{\ds}{\displaystyle}
\newcommand{\pari}{{\scriptscriptstyle\parallel}}
\newcommand{\goo}{\,\raisebox{-.5ex}{$\stackrel{>}{\scriptstyle\sim}$}\,}
\newcommand{\loo}{\,\raisebox{-.5ex}{$\stackrel{<}{\scriptstyle\sim}$}\,}
\newcommand{\Gt}{\Gamma_{\omega^{\scriptstyle *}}}
\newcommand{\Gti}{\Gamma_{\mbox{$\scriptstyle\vom\to i$}}}

\begin{document}

\baselineskip 24pt
\begin{center}
%---------------------------------------------------------
{\Large\bf Dilepton production by bremsstrahlung of\\
meson fields in nuclear collisions}\\[5mm]
%---------------------------------------------------------
{\bf I.N.~Mishustin$^{a,b}$, L.M.~Satarov$^a$, H.~St\"ocker$^c$ and
W.~Greiner$^c$}
\end{center}
\baselineskip 16pt
\begin{tabbing}
\hspace*{4em}\=${}^a$\,\={\it The Kurchatov~Institute,
123182~Moscow,~\mbox{Russia}}\\
\>${}^b$\>{\it The Niels~Bohr~Institute,
DK--2100~Copenhagen {\O},~\mbox{Denmark}}\\
\>${}^c$\>{\it Institut~f\"{u}r~Theoretische~Physik,
J.W.~Goethe~Universit\"{a}t,}\\
\>\>{\it D--60054~Frankfurt~am~Main,~\mbox{Germany}}
\end{tabbing}
\baselineskip 24pt

\begin{abstract}
We study the bremsstrahlung of virtual $\omega$--mesons
due to the collective deceleration of nuclei at the initial stage of
an ultrarelativistic heavy--ion collision. It is shown that
electromagnetic decays of these mesons may give an important
contribution to the observed yields of dileptons. Mass spectra of
$e^+e^-$ and $\mu^+\mu^-$ pairs produced in central Au+Au collisions
are calculated under some simplifying assumptions on the space--time
variation of the baryonic current in a nuclear collision process.
Comparison with the CERES data for 160 AGev Pb+Au collisions shows that
the proposed mechanism gives a noticeable fraction of the observed
$e^+e^-$ pairs in the intermediate region of invariant masses.
Sensitivity of the dilepton yield to the in--medium modification of
masses and widths of vector mesons is demonstrated.
\end{abstract}

According to the relativistic mean--field model~\cite{Wal85}, strong
time--dependent meson fields are generated in the course of a relativistic
heavy--ion collision. Using the approach developed in papers on the
pion~\cite{Vas80} and photon~\cite{Lip88} bremsstrahlung
we suggested recently~\cite{Mis95} a new mechanism of particle production
by the collective bremsstrahlung and decay of classical meson fields
in relativistic heavy--ion collisions. The comparison with the observed
data on pion multiplicity shows \cite{Mis96} that this mechanism may be
important in central collisions of most heavy nuclei already at the SPS
bombarding energy.

Within this mechanism the production of some particle(s) $i$ is
considered as a two--step process \mbox{$A_p+A_t\to\vom\to
\mbox{$i+X$}$}. Here $A_p (A_t)$ stands for the projectile (target)
nucleus and $\vom$ is a virtual vector meson
\footnote{Here and below the virtual particles ($\omega$ and $\gamma$)
are marked by a superscript `\raisebox{-.5ex}{*}'.}.
The first step in the
above reaction corresponds to the virtual bremsstrahlung process
leading to the creation of an off--mass--shell vector meson. The
second step is the superposition of all channels of the virtual meson
decay with the particle~$i$ in the final state. Below we consider the
production of virtual mesons in the coherent process caused by the
collective deceleration of the projectile and target nuclei at the
initial stage of the reaction. Preliminary results concerning the
contribution of the above mechanism to the production of pions,
real $\omega$--mesons and dileptons are published in
Ref.~\cite{Mis96}.

Below we focus mainly on the dilepton emission in central
collisions of ultrarelativistic nuclei. The analysis of the dilepton
production is interesting at least by two reasons. First, dileptons are
highly penetrating particles and carry practically an undistorted
information about their creation points. Second, a strong enhancement
of the dilepton yield was observed recently in central 200 AGeV
S+Au~\cite{Cer95}, S+W~\cite{Hel95} and 160 AGeV Pb+Au~\cite{Cer96}
% 9,10,11: \cite{Li95,Cas95,UQM96}
collisions. The analysis shows~[9--11] that this enhancement can only
partially be
explained by the conventional mechanism of binary hadron collisions,
e.g. by \mbox{$\pi\pi\to\rho\to{\rm e}^+{\rm e}^-$} processes.
According to Refs.~\cite{Li95,Cas95} the experimental data can be
reproduced assuming a strong reduction of the $\rho$ meson mass in
dense and hot nuclear matter.  On the other hand, as argued in
Ref.~\cite{UQM96}, the in--medium effect was probably overestimated in
these calculations.
%The inclusive Pb + Au data can be reproduced by
%the UQMD model \cite{UQM96} without the $\rho$ mass shift.
Below we show that the enhanced dilepton yield
may be explained, at least partly, by the contribution of the
collective bremsstrahlung mechanism.

\section*{1.~Formulation of the model}

\hspace{1.5em}By analogy to the Walecka model we introduce the
vector meson field $\omega^\mu(x)$ coupled to the 4--current $J^\mu(x)$
of baryons participating in a heavy--ion collision at a given impact
parameter~$b$. The equation of motion determining the space--time
behaviour of $\omega^\mu(x)$ can be written as (\mbox{$c=\hbar=1$})
\bel{eqm}
(\partial^\nu\partial_\nu+m_\omega^2)\,\omega^\mu(x)=g_V J^\mu(x)\,,
\ee
where $g_V$ is the $\omega NN$ coupling constant and
$m_\omega\simeq$ 783 MeV is the omega meson mass. In the mean--field
approximation $\omega^\mu$ is considered as a purely classical field.
From~\re{eqm} one can
see that excitation of propagating waves in the vacuum (the
bremsstrahlung process) is possible if the Fourier transformed baryonic
current
\bel{ftbc} J^\mu(p)=\int{\rm d}^4x J^\mu(x) e^{ipx}
\ee
is nonzero in the time--like region $p^2\sim m_\omega^2$.

In the following we study the bremsstrahlung process in the lowest order
approximation neglecting the back reaction and reabsorption of the emitted
vector mesons, i.e. treating $J^\mu$ as an external current. From~\re{eqm}
one can calculate the energy flux of the vector field at a large distance
from the collision region~\cite{Vas80}. Then this flux is expressed
in terms of the distribution of field quanta, i.e. $\omega$--mesons.
This leads to the following
formulae for the momentum distribution of real $\omega$--mesons
emitted in a heavy--ion collision~\cite{Iva89}
\bel{rvms}
E_\omega\frac{\ds{\rm d}^3 N_\omega}
{\ds{\rm d}^3p\hfill}=S(E_\omega,\bm{p})\,,
\ee
where $E_\omega=\sqrt{m_\omega^2+\bm{p}^{\,2}}$ and
\bel{sfu}
S(p)=\frac{g_V^2}{16\pi^3}|J_\mu^*(p)J^\mu(p)|
\ee
is a source function. In our model the latter is fully determined
by the collective motion of the projectile and target nucleons.

To take into account the off--mass--shell effects we characterize
virtual $\omega$ mesons by the mass $M_\omega\equiv\sqrt{p^2}$
and total width $\Gamma_{\vom}$. The spectral function
of virtual $\omega$ mesons may be written as
\bel{spf}
\rho(M_\omega)=\frac{2}{\pi}\frac{\ds M_\omega\Gt}
{\ds (M_\omega^2-m_\omega^2)^2+m_\omega^2\Gt^2}\,.
\ee
%Below virtual particles ($\omega$ and $\gamma$) are marked
%by the superscripts '\raisebox{-.5ex}{*}'.
To calculate the distribution of virtual mesons in their 4--momenta
$p$ we use the formulae~\cite{Mis96}
\bel{dvom}
\frac{\ds{\rm d}^4 N_{\vom}}
{\ds{\rm d}^4p\hfill}= \rho(M_\omega) S(p)\,.
\ee
In the limit \mbox{$\Gamma_{\scriptstyle\vom}\to 0$} one can replace
$\rho(M_\omega)$ by $2\delta(M_{\omega}^2-m_\omega^2)$\,. In this
case~\re{dvom} becomes equivalent to the formulae~(\ref{rvms}) for the
spectrum of the on--mass--shell vector mesons. Below we study also
how the dilepton production is changed when the pole
position in the vector meson propagator is shifted due to the medium
effects.

We consider the following channels of the virtual $\omega$ decay,
most important at invariant masses $M_{\omega}\loo m_\omega$:
$i={\ds e}^+{\ds e}^-$, $\mu^+\mu^-$, $\pi^0\gamma$,
$\pi^0{\ds e}^+{\ds e}^-$, $\pi^0\mu^+\mu^-$, $\pi^+\pi^-$,
$\pi^+\pi^-\pi^0$\,. The total width $\Gt$ is decomposed into the
sum of partial decay widths $\Gti$\,:
\bel{pdw}
\Gt=\sum_i{\Gti}\,.
\ee
The distribution over the total 4--momentum of particles in a given
decay channel can be written as
\bel{dtmi}
\frac{\ds{\rm d}^4 N_{\mbs{\vom\to i}}}
{\ds{\rm d}^4p\hfill}=B_{\mbs{\vom\to i}}
\frac{\ds{\rm d}^4 N_{\vom}}{\ds{\rm d}^4p\hfill}\,,
\ee
where $B_{\mbs{\vom\to i}}\equiv\Gti/\Gt$ is the
branching ratio of the $i$--th decay channel. The latter is a function
of the total invariant mass of the decay
particles~$M=\sqrt{p^2}=M_\omega$.

To calculate the 4--current $J^\mu(p)$ determining the source function
$S(p)$ we adopt the simple picture of a high--energy heavy--ion
collision suggested in Ref.~\cite{Mis95}. We consider collisions
of identical nuclei ($A_p=A_t=A$) at zero impact parameter.  In the
equal velocity frame the projectile and target nuclei initially move
towards each other with velocities $\pm v_0$ or rapidities $\pm y_0$,
where $v_0=\tanh{y_0}=(1-4m_N^2/s)^{1/2}$ and $\sqrt{s}$ is the c.m.
bombarding energy per nucleon. In the ''frozen density''
approximation~\cite{Mis95} the internal compression and transverse
motion of nuclear matter are disregarded at the early
(interpenetration) stage of the reaction. Within this approximation the
colliding nuclei move as a whole along the beam axis with instantaneous
velocities $\dot{z}_p=-\dot{z}_t\equiv\dot{z}(t)$\,. The projectile
velocity $\dot{z}(t)$ is a decreasing function of time, which we
parametrize in the form~\cite{Vas80}
\bel{prtr}
\dot{z}(t)=v_f+\frac{\ds v_0-v_f}{\ds 1+{\rm e}^{t/\tau}}\,,
\ee
where $\tau$ is the effective deceleration time
and $v_f$ is the final velocity of nuclei (at $t\to +\infty$)\,.

In this approximation the Fourier transform of the baryon current
 $J^\mu(p)$ is totally
determined by the projectile trajectory $z(t)$~\cite{Mis95}:
\bel{ftre}
J^0(p)=\frac{p_\pari}{p_0}J^3(p)=
2A\int\limits_{-\infty}^\infty{\rm d}t
e^{\ds ip_0t}\cos{[p_\pari z(t)]}\,F\left(\sqrt{\bm{p}_T^2+
p_\pari^2\cdot[1-\dot{z}^2(t)]}\right)~,
\ee
where $p_\pari$ and $\bm{p}_T$ are, respectively, the longitudinal
and transverse components of the three--momentum $\bm{p}$,  $F(q)$ is
the density form factor of the initial nuclei
\bel{dff}
F(q)\equiv\frac{1}{A}\int{\rm d}^3r\rho(r) e^{\ds-i\bm{q}\,\bm{r}}
=\frac{4\pi}{Aq}\int\limits_{0}^{\infty}r{\rm d}r\rho(r)\sin{qr}\,.
\ee
The time integrals in \re{ftre} were calculated numerically assuming
the Woods--Saxon distribution of the nuclear density $\rho(r)$\,.

In this work we choose the same coupling constant, $g_V$=13.78 and
stopping parameters $\tau, v_f$ as in \mbox{Ref.~\cite{Mis95}}. In
particular, it is assumed that $\tau$ equals one half of the nuclear
passage time
\bel{tauc}
\tau=\frac{\ds R}{\ds\sinh{y_0}}\,,
\ee
where $R$ is the geometrical radius of initial nuclei, $R=r_0 A^{1/3}$
with $r_0=1.12$ fm. Instead of $v_f$ it is more convenient
to introduce the c.m. rapidity loss $\delta y$ defined by the relation:
\bel{vfin}
v_f=\tanh{(y_0-\delta y)}\,.
\ee
For central Au+Au collisions we assume the energy--independent
value~\cite{Sch93} $\delta y=2.4$ for $\sqrt{s}>10$ GeV and the full
stopping ($\delta y=y_0$) for lower bombarding energies.

\section*{2.~Different channels of virtual meson decay
and dilepton production}

\hspace{1.5em}Similarly to Ref.~\cite{Koc93} we assume that the ''direct''
decays of $\omega$ mesons into dileptons, \mbox{$\vom\to l^+l^-$},
proceed via the intermediate emission and decay of virtual photons
$\gamma^{\scriptstyle *}$. In the following the explicit formulae are
presented for the $e^+e^-$ production only.  The corresponding
expressions for dimuons are obtained by replacing the lepton
masses and decay widths. The matrix element of the process
\mbox{$\vom\to e^+e^-$} is proportional to the $\omega$ meson
polarization vector $\xi_\mu$, the lepton current
$\overline{v}_+\gamma^\mu u_-$ and the photon propagator $k^{-2}$, where
$k=p_++p_-$ is the total 4--momentum of the lepton pair. The
calculation of the partial decay width gives the result~\cite{Koc93}
\bel{wdl}
\Gamma_{\mbs{\vom\to ee}}(M)\propto
M^{-3}\,\Gamma_{\mbs{\gamma^{\scriptstyle *}\to ee}}\,.
\ee
Here $M=\sqrt{k^2}$ is the dilepton invariant mass (in the
direct channel $M=M_\omega$) and
$\Gamma_{\mbs{\gamma^{\scriptstyle *}\to ee}}$
is the partial width of a virtual photon:
\bel{gdl}
\Gamma_{\mbs{\gamma^{\scriptstyle *}\to ee}}=
\frac{\ds\alpha\beta}{\ds 2}\left(1-\frac{\ds\beta^2}{\ds 3}\right)
\Theta(M-2m_e)\,,
\ee
where $\alpha=e^2/\hbar c$,
$\Theta(x)\equiv\frac{1}{2}(1+{\rm sign}{x})$, $m_e$
is the electron mass and
\bel{bet}
\beta=\sqrt{1-\frac{\ds 4m_e^2}{\ds M^2}}\,.
\ee
The proportionality coefficient in \re{wdl} is determined from the
condition $B_{\mbs{\vom\to ee}}(m_\omega)=B_{ee}$ where
$B_{ee}=7.1\cdot 10^{-5}$ is the observed probability of the
\mbox{$\omega\to ee$} decay~\cite{PDG94}.

We take into account also the three-particle, ''Dalitz''
decays \mbox{$\vom\to\pi^0 e^+e^-$}.
At fixed values of $M$ and $M_\omega$ the components of the total
dilepton 4-momentum $k$ in the $\omega$ rest frame can be found
by using the expressions
\bel{pmom}
k_0=\sqrt{\bm{k}^2+M^2}=\frac{M^2+M^2_\omega-m^2_\pi}{2M_\omega}\,,
\ee
where $m_\pi=0.14$ GeV is the pion mass.
The corresponding partial width can be calculated assuming that the
Dalitz decay is the two--step process
\mbox{$\vom\to\pi\gamma^*\to\pi ee$}. Generalizing the
results of~\cite{Lan85} to the case of virtual $\omega$'s we obtain the
following expression for the differential width of the Dalitz decay
\bel{dald}
\frac{\ds {\rm d}\Gamma_{\mbs{\vom\to\pi ee}}}{\ds {\rm d}M\hfill}=
\frac{\ds 2}{\ds\pi M^2}\Gamma_{\mbs{\vom\to\pi\gamma^*}}
\Gamma_{\mbs{\gamma^{\scriptstyle *}\to ee}}\,.
\ee
The \mbox{$\omega^{\scriptstyle *}\to\pi\gamma^*$} decay width is
proportional to the electromagnetic form factor squared,
$F^2_{\omega\pi}$:
\bel{wpg}
\Gamma_{\mbs{\vom\to\pi\gamma^*}}\propto
F^2_{\omega\pi}|\bm{k}|^3\Theta(M_\omega-M-m_\pi)\,,
\ee
where $|\bm{k}|$ is determined from \re{pmom}. The coefficient
of proportionality may be found by considering the limiting case
\mbox{$M_\omega\to m_\omega$}, \mbox{$M\to 0$}, when the
left hand side of Eq. (\ref{wpg}) coincides with the observed width of
the \mbox{$\omega\to\pi\gamma$} decay. As shown in~\cite{Lan85}
the experimental data for $F_{\omega\pi}$ are well reproduced within
the vector meson dominance model~\cite{Sac69}. Assuming that this model
is valid also for decays of virtual $\omega$'s we have
\bel{pfor}
F_{\omega\pi}^2=\frac{\ds m_\rho^2\,(m_\rho^2+\Gamma^2_\rho)}
{\ds (M^2-m^2_\rho)^2+m_\rho^2\Gamma^2_\rho}\,,
\ee
where $m_\rho$ and $\Gamma_\rho$ are the mass and total width of the
$\rho$ meson. Unless otherwise stated, \re{pfor} is used  with the
parametrization $\Gamma_\rho=\Gamma_\rho(M)$ suggested in
Ref.~\cite{Koc93} and the free $\rho$ meson mass
($m_\rho=m_{\rho 0}\simeq$~0.77 GeV).

To calculate the total width of virtual $\omega$'s
one should also know the partial widths of non-electromagnetic
decay channels. In the considered region of masses
$M_\omega\loo$~1 GeV we take into account the decays with two and three
pions in the final state. Assuming that the
\mbox{$\omega^{\scriptstyle *}\to 2\pi$} matrix
element is proportional to the product of the $\omega$
meson polarization vector and the difference of the pion 4--momenta, we get
\bel{twop}
\Gamma_{\mbs{\vom\to 2\pi}}(M_\omega)\propto
M_\omega^{-2} \cdot\left(M_\omega^2-4m_\pi^2\right)^{3/2}
\Theta(M_\omega-2m_\pi)\,.
\ee
The proportionality coefficient is taken from the condition
$B_{\mbs{\vom\to 2\pi}}(m_\omega)=B_{2\pi}=0.022$~\cite{PDG94}.
The \mbox{$\vom\to 3\pi$} partial width is calculated assuming
that it is proportional to the 3--pion phase space
volume allowed by the kinematics~\cite{Iva89}:
\bel{thrp}
\Gamma_{\mbs{\vom\to 3\pi}}(M_\omega)\propto
\Phi_{3\pi}(M_\omega)
\cdot\Theta(M_\omega-3m_\pi)\,,
\ee
with the coefficient determined from the relation
$B_{\mbs{\vom\to 3\pi}}(m_\omega)=B_{3\pi}=0.89$~\cite{PDG94}.

The mass distribution of $e^+e^-$ pairs produced by the
bremsstrahlung mechanism can be written as a convolution
of the virtual $\omega$ meson spectrum and the differential branching
of the \mbox{$\vom\to eeX$} decay  ($X$ denotes any particle(s)
emitted together with the lepton pair):
\bel{mdil}
\frac{\ds{\rm d}N_{ee}}{\ds{\rm d}M\hfill}=\int{\rm d}^4p_\omega
\frac{\ds{\rm d}^4N_{\omega_*}}{\ds{\rm d}^4p_\omega}\cdot
\frac{\ds{\rm d}B_{\mbs{\vom\to eeX}}}{\ds{\rm d}M\hfill}\,.
\ee
Taking into account only the direct and Dalitz decays,
the dilepton mass distribution can be represented as:
\bel{mdil1}
\frac{\ds{\rm d}N_{ee}}{\ds{\rm d}M\hfill}=
B_{\mbs{\vom\to ee}}\frac{\ds{\rm d}N_{\omega_*}}{\ds{\rm
d}M\hfill}+\int\limits_{M+m_\pi}^\infty{\rm d}M_\omega
\frac{\ds{\rm d}N_{\omega_*}}{\ds{\rm d}M_\omega\hfill}\cdot
\frac{\ds{\rm d}B_{\mbs{\vom\to ee\pi}}}{\ds{\rm d}M\hfill}\,.
\ee
The first (direct) term is obtained by using the relation
${\rm d}B_{\mbs{\vom\to ee}}/{\rm d}M=
B_{\mbs{\vom\to ee}}\delta(M-M_\omega)$ and performing the explicit
integration over $M_\omega$.

To compare the model predictions with experimental data one should
take into account the various acceptance cuts used in different experiments.
This severely complicates the calculations: in general
one should know the differential branching
${\rm d}^6B{\mbs{\vom\to eeX}}/{\rm d}^3p_+{\rm d}^3p_-$ which describes
the probability of the $\omega$ meson decay into the lepton pair with
the positron and electron \mbox{3--momenta} $\bm{p}_+$ and $\bm{p}_-$,
respectively. To calculate the acceptance weighted mass distribution
one should replace ${\rm d}B_{\mbs{\vom\to eeX}}/{\rm d}M$ in \re{mdil} by
\bel{mdil2}
\frac{\ds{\rm d}B^{(A)}_{\mbs{\vom\to eeX}}}{\ds{\rm d}M\hfill}=
\int {\rm d}^3p_+{\rm d}^3p_-\,{\cal A}\cdot\frac{\ds{\rm d}^6B_{\mbs
{\vom\to eeX}}} {\ds{\rm d}^3p_+{\rm d}^3p_-\hfill}
\delta(M-\sqrt{k^2})\,.
\ee
The weight function ${\cal A}$ equals one (zero) if
$\bm{p}_\pm$ are inside (outside) the kinematical volume covered
in a given experiment. In numerical calculations presented in the next
section we use for $e^+e^-$ pairs the acceptance cuts of the CERES
experiment~\cite{Cer96}
\bel{acv}
p_{T\pm} > 0.175\ {\rm GeV/c},\ 2.1< \eta_\pm < 2.65, \
\theta_{ee} > 0.035\,.
\ee
Here $p_{T\pm}$ and $\eta_\pm$ are the transverse momenta and
pseudorapidities of leptons, $\theta_{ee}$ is their relative emission
angle in the lab frame.

At given $M$ and $M_\omega$ the components of the vectors $\bm{p}_\pm$
are fixed by the angular variables \mbox{$\Omega=(\theta,\phi)$}
and $\td{\Omega}=(\td{\theta},\td{\phi})$, where
$\Omega$ denotes spherical angles of $\bm{k}$ with respect to
the $\omega$ meson 3--momentum (in its rest frame) and $\td{\Omega}$
stands for the positron emission angles with respect to $\bm{k}$ (in
the pair rest frame). By using these variables one can rewrite
\re{mdil2} as follows
\bel{mdil3}
\frac{\ds{\rm d}B^{(A)}_{\mbs{\vom\to eeX}}}{\ds{\rm d}M\hfill}=
\frac{\ds{\rm d}B_{\mbs{\vom\to eeX}}}{\ds{\rm d}M\hfill}
\mbox{${\ds\int}{\rm d}\Omega{\ds\int}{\rm d}\td{\Omega}$}\,
\frac{\ds{\rm d}W_{\mbs{\vom\to eeX}}}
{\ds{\rm d}\Omega{\rm d}\td{\Omega}\hfill}\,{\cal A}\,,
\ee
where ${\rm d}W_{\mbs{\vom\to eeX}}/{\rm d}\Omega{\rm d}\td{\Omega}$
is the angular distribution of the \mbox{$\vom\to eeX$} decay
normalized to unity.

In our case the usual procedure of calculating the direct and Dalitz
parts of dilepton distributions by simple averaging over all
polarizations of decaying $\omega$ mesons is not correct. Indeed, as
seen from \re{eqm}, the polarization vector of a virtual vector meson
$\xi_\mu$ is proportional to $J_\mu(p)$ where $p$ is the meson
4--momentum. In the $\omega$ meson rest frame $\xi_\mu=(0,\bm{\xi})$,
where \bm{\xi} is a vector parallel to the direction of \bm{p}. Proceeding
from the \mbox{$\vom\to ee$} matrix element (see above) we get the
following relation for the direct part of the angular distribution
\bel{dira}
\frac{\ds{\rm d}W_{\mbs{\vom\to ee}}}{\ds{\rm d}
\Omega{\rm d}\td{\Omega}\hfill}=C_{\rm dir}\,
\left(1-\beta^2\cos^2{\td{\theta}}\right)\,\delta(\Omega)\,,
\ee
where $C_{\rm dir}$ is the normalization constant. The anisotropy of the
lepton angular distribution in the rest frame of the $\omega$ meson is
a consequence of its polarization.

The Dalitz part of the dilepton distribution is calculated
assuming~\cite{Koc93,Lan85} that the \mbox{$\vom\to ee\pi$} matrix
element is proportional to $\epsilon_{\mu\nu\sigma\delta}\xi^\mu p^\nu
k^\sigma (\overline{v}_+\gamma^\delta u_-)$\,. The direct calculation gives
\bel{Dala}
\frac{\ds{\rm d}W_{\mbs{\vom\to ee\pi}}}{\ds{\rm d}
\Omega{\rm d}\td{\Omega}\hfill}=C_{\rm dal}\,
\sin^2{\theta}\left[1-\beta^2\sin^2{\td{\theta}}\,
\sin^2(\td{\phi}-\phi)\right]\,,
\ee
where $C_{\rm dal}$ is found from the normalization condition.
The averaging over the $\omega$ meson polarizations is equivalent
to the averaging over $\Omega$. As a result we obtain the distribution
over $\td{\Omega}$ obtained earlier in Ref.~\cite{Kro55}.

\section*{3.~Results}

\hspace{1.5em}Let us now discuss the results of numerical calculations
obtained within the model described above. One should bear in
mind that the model assumes a rather simplified space---time evolution,
in particular the collective
projectile--target deceleration (see the discussion in
Ref.~\cite{Mis95}). Therefore, the model in its present form can
be used for a qualitative analysis only.

Fig.~1 shows the dilepton mass spectrum in central ($b=0$) 160 AGeV
Au+Au collisions. One can see that the mass distribution of dileptons
produced by the virtual $\omega$ decays is quite different from that
predicted by the conventional hadronic sources. In particular, the low
and intermediate mass region is strongly enhanced. This is explained by
the copious production of ``soft'' virtual $\omega$'s by the bremsstrahlung
mechanism. The contribution of direct $\omega$ decays is peaked at
very small invariant masses as well as at the pole position of the $\omega$
propagator, $M=m_{\omega}$. The Dalitz
contribution is most important in the intermediate region of dilepton
masses, 0.2 GeV $\loo M\loo$ 0.6 GeV.
A similar behaviour is predicted for the dimuon spectrum
\footnote{Since the branching ratio of the direct decay
\mbox{$\omega\to\mu^+\mu^-$} is not known experimentally~\cite{PDG94}
we assume that it is equal to $B(\omega\to e^+e^-)$.},
Fig.~2. The main difference here is the much higher mass
threshold at $M=2m_\mu$.

In Fig.~3 we compare the model predictions for the same reaction, but
at different bombarding energies, $\sqrt{s}=17.43$ AGeV (SPS) and 200 AGeV
(RHIC). We have also calculated the dilepton spectra at
the LHC energy $\sqrt{s}$=6.3 ATeV but the corresponding results
practically coincide with the model prediction for the RHIC energy.
Such a behavior follows from the energy independence of the stopping
parameter $\delta y$, assumed at high $\sqrt{s}$ (see Sect.~1). As a
consequence, the phase--space distribution of primordial $\omega$
mesons, produced by the bremsstrahlung mechanism, saturates with raising
bombarding energy~\cite{Mis95}.

%To compare the model results with experimental data it is necessary to
%take into account
The experimental acceptance cuts and a poor mass resolution
distort significantly the dilepton mass distributions as compared to
those presented in Figs.~1--3. In Fig.~4 we show the dilepton mass
spectrum for central 160 AGeV Au+Au collisions. The CERES
acceptance cuts and mass resolution are included in this calculation.
The double differential spectrum $d^2N_{ee}/dMd\eta$ is obtained by
dividing the acceptance weighted mass distribution, \re{mdil1}, by the
width of the CERES pseudorapidity window. In calculating the
acceptance weight $\cal{A}$ entering \re{mdil3} we have neglected the
transverse momenta of primordial $\omega$ mesons (see
Ref.~\cite{Mis95}). At the same figure we show separately the
contributions of direct and Dalitz decays of $\omega$ mesons.  Note
that the original spectrum (without mass resolution corrections) of
e$^+$e$^-$ pairs has a strong peak at $M\simeq m_\omega$. This peak
originates from the direct $\omega$ meson decays. The step--like
behaviour of the direct contribution at $M\simeq 0.35$ GeV appears due
to the CERES cut at small transverse momenta,
\mbox{$p_{T\pm}>p_{\rm min}$=0.175 GeV/c}. Indeed, in the limit
\mbox{$p_{T\omega}=0$} the minimal invariant mass of ''direct''
pairs is $2\,\sqrt{m_e^2+p_{\rm min}^2}\approx$ 0.35 GeV. Taking into
account nonzero components of $\bm{p}_{T\omega}$ will result in a certain
smoothening of the above--mentioned jump in the dilepton mass distribution.

As one can see from Fig.~4, the bremsstrahlung mechanism gives
a significant contribution to the dilepton production in
the intermediate mass region. Hopefully, this contribution can be
observed experimentally by using a characteristic
angular distribution in the dilepton momentum predicted by
the model~\cite{Mis95,Mis96}. However, the bremsstrahlung
contribution alone is not sufficient to explain the dilepton yield
observed in central 160 AGeV Pb+Au collisions. In the most interesting
region of masses \mbox{$M\simeq 0.4-0.6$ GeV} the data are
underestimated by a factor of about three.

A special calculation showed that this discrepancy can not be removed
by taking into account the excitation and decays of virtual $\rho$
mesons disregarded in the present calculation. The contribution of the
$\rho$ meson bremsstrahlung in symmetric heavy--ion collisions is
proportional to the isospin asymmetry factor $\chi=(1/2-Z/A)^2$ where
$Z$ and $A$ are the charge and mass numbers of the colliding nuclei.
Since $\chi\loo 10^{-2}$, the role of the $\rho$ meson bremsstrahlung
is relatively small even for heaviest nuclei.

Of course, in addition to the collective bremsstrahlung mechanism the usual,
incoherent, sources of dilepton production (e.g. the $\pi\pi\to ee$ and
$\pi\to ee\gamma$ processes) also give a noticeable contributions.
% 9,10,11 \cite{Li95,Cas95,UQM96}
According to Refs.~[9--11] the dynamical models incorporating only these
incoherent hadronic sources may easily explain the low mass dilepton
yields in S+Au and Pb+Au collisions at the SPS energies. On the other
hand, these models significantly underestimate the observed data in the
intermediate mass region. As argued in Refs.~\cite{Li95,Cas95} the
agreement with experimental data can be achieved if one assumes a
strong reduction of the vector meson masses in dense nuclear matter.
%However, this conclusion is not supported by other authors~\cite{UQM96}.

To check the sensitivity of our model to the in--medium modification of
the vector mesons, in \mbox{Fig.~5} we compare the dilepton mass
distributions calculated for different values of the $\rho$
and $\omega$ masses. To diminish the number of model parameters we take
%assume the Brown--Rho scaling~\cite{Br91} taking
the same mass reduction factor for $\rho$ and $\omega$ mesons:
\mbox{$m_\rho/m_{\rho 0}=m_\omega/m_{\omega 0}\equiv R_m$\,.} One can see
that the dilepton mass distribution is rather sensitive to $R_m$.
A relatively small, 20~\%,~reduction of the meson masses raises
the dilepton yield at $M\sim$ 0.5 GeV by a factor of about two.
On the other hand, the model calculation with fixed $R_m$ predicts too
high peaks in mass distributions. One should bear in mind that these
calculations provide only a rough estimate of the possible effect since in
an actual nuclear collision mass shifts should be time (and
space) dependent. Therefore, the observed distribution will be a
superposition of contributions with different $R_m$\,. As a result,
the peak of direct dileptons will be less pronounced in this
distribution.

%misspell: df->of
As calculations of Ref.~\cite{Wo96} show, the $\omega$ meson width may
be significantly increased in dense baryonic matter due to the mixing of
$\sigma$ and $\omega$ mesons~\cite{Wal85}. At baryonic densities
of about two times the normal nuclear density, the width may
increase by a factor of about 5 as compared to its vacuum value.
To estimate the effects of the in--medium broadening of virtual $\omega$
mesons, we have performed the calculation where the partial decay
widths were scaled by the same
amplification factor $\lambda$, independent of $M_\omega$. The
calculation shows that this broadening influences mainly the direct
component of the dilepton yield. As seen from Fig.~6, the yield
%agreement with the observed data
in the intermediate mass region may be strongly enhanced
at $\lambda\goo 5$.

\section*{4.~Summary}

\hspace{1.5em}In conclusion, we have shown that the collective
bremsstrahlung of the vector meson field can provide an important source of
dilepton production in high energy heavy--ion collisions. This
mechanism may be responsible, at least partly, for the enhanced yield
of dileptons observed in central nuclear collisions at the SPS
bombarding energies. It has been demonstrated that the coherent dilepton
production is sensitive to the in--medium modification of vector
meson masses and widths. Obviously, these effects should be
studied in more details in microscopic models.

%To make the model more realistic,
In future studies one should implement a more
realistic dynamical picture of a heavy--ion collision
by using either the fluid--dynamical or kinetic approaches. In this
way one can take into account the flow and compression effects
disregarded in present model. Also the formalism should be
generalized to study bremsstrahlung effects in a situation when masses
and widths of vector mesons are space and time dependent. To extend
the calculations to collider energies, $\sqrt{s}\geq200$ GeV,
the model should be reformulated at the quark--gluon level.

\section*{5.~Acknowledgments}

\hspace{1.5em}The authors thank J.P. Bondorf and L.A. Winckelmann for
useful discussions, J. Stachel and Th.~Ullrich for providing a valuable
information on CERES data. This work has been supported in part by the
Carlsberg Foundation (Denmark) and EU--INTAS Grant No.~94--3405.
We acknowledge also the financial support from GSI, BMFT and DFG.
Two of us (I.N.M., L.M.S.) thank the
Institute for Theoretical Physics, University of Frankfurt am Main,
and the Niels Bohr Institute, University of Copenhagen,
for the kind hospitality and financial support.

\vspace{0.5cm}
%\newpage

\newpage

%------------------------------------------------------
\section *{Figure captions}
%------------------------------------------------------
\newcommand{\Fig}[2]{\noindent{\bf Fig.~#1.~}
\parbox[t]{15cm}{\baselineskip 24pt #2}\\[5mm]}
\Fig{1}
{Mass spectrum of $e^+e^-$ pairs
produced by the collective brems\-strahlung mechanism in central
160~AGeV Au+Au collisions. Contributions of the direct
(\mbox{$\omega_*\to ee$}) and Dalitz (\mbox{$\vom\to\pi ee$}) decay
channels are shown by dotted and dashed line respectively.}
\Fig{2}
{The same as in Fig.~1, but for the spectrum of $\mu^+\mu^-$ pairs.}
\Fig{3}
{Comparison of $e^+e^-$ yields
in central Au+Au collisions at the SPS and RHIC bombarding energies.}
\Fig{4}
{Mass spectrum of $e^+e^-$ pairs produced by the collective bremsstrahlung
mechanism
in central 160 AGeV Au+Au collisions. The grey histogram (solid line) shows
the model results with (without) inclusion of experimental
mass resolution. The dotted (dashed) line shows the contribution of
direct (Dalitz) $\omega$ meson decays. Preliminary
experimental data for central Pb+Au collisions~\cite{Cer96}
are shown by solid circles.}
\Fig{5}
{Comparison of $e^+e^-$ spectra in central 160 AGeV Au+Au
collision for different values of the meson mass reduction
factor $R_m$. Preliminary experimental data for central Pb+Au collisions
are taken from Ref.~\cite{Cer96}.}
\Fig{6}
{The same as in Fig.~5 but for different choices
of the virtual $\omega$ width in units of the vacuum width $\Gamma_0=8.4$ MeV.}
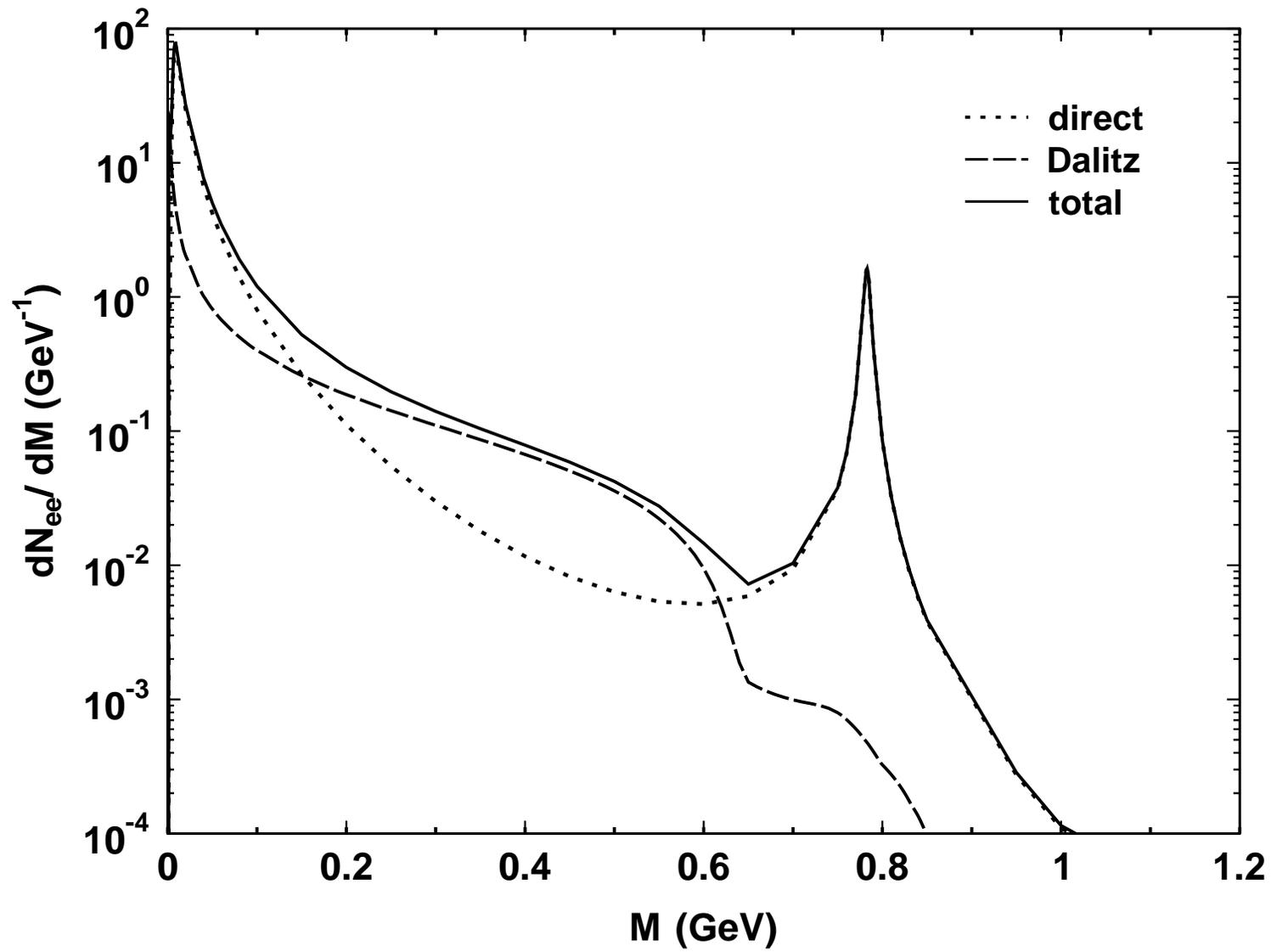
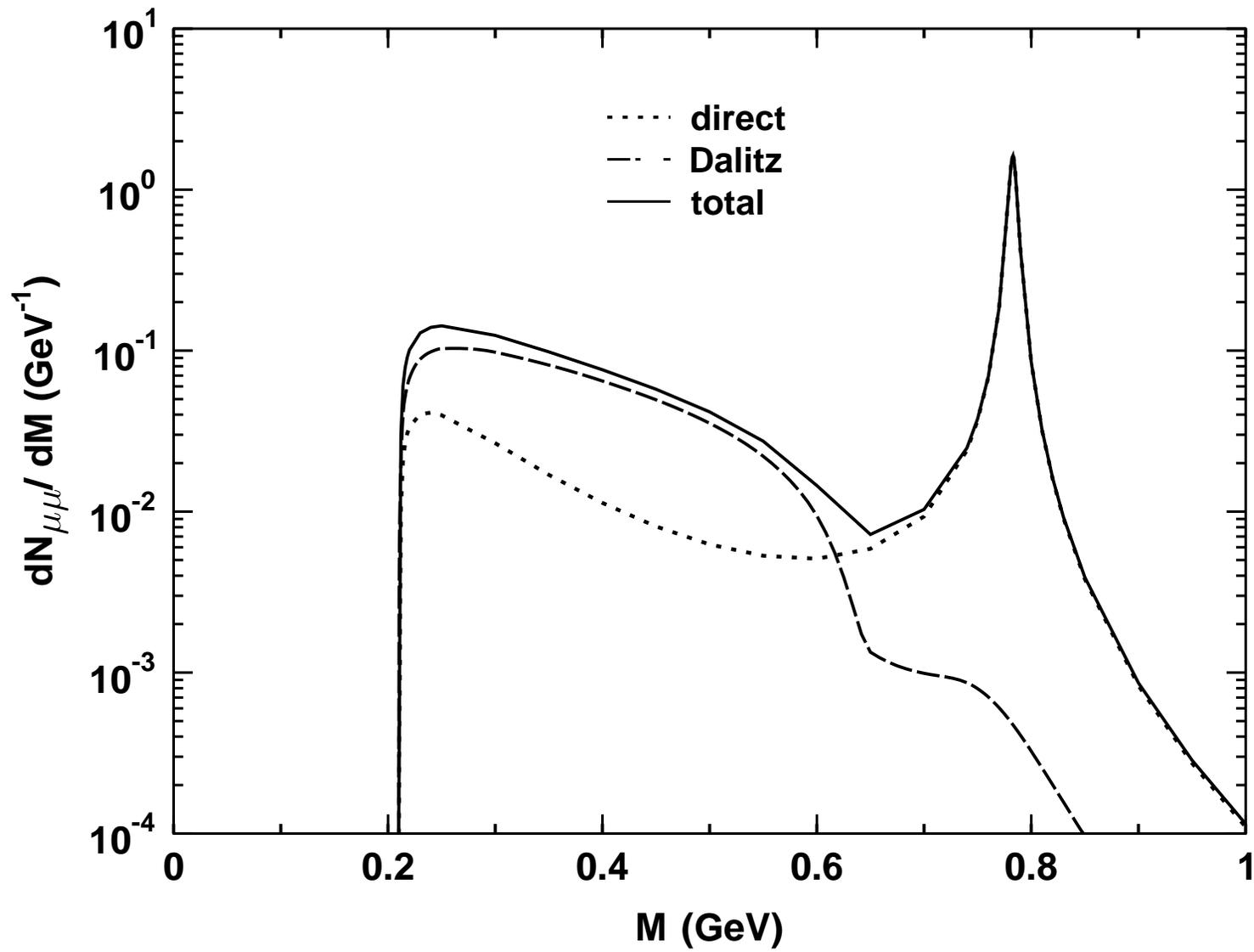
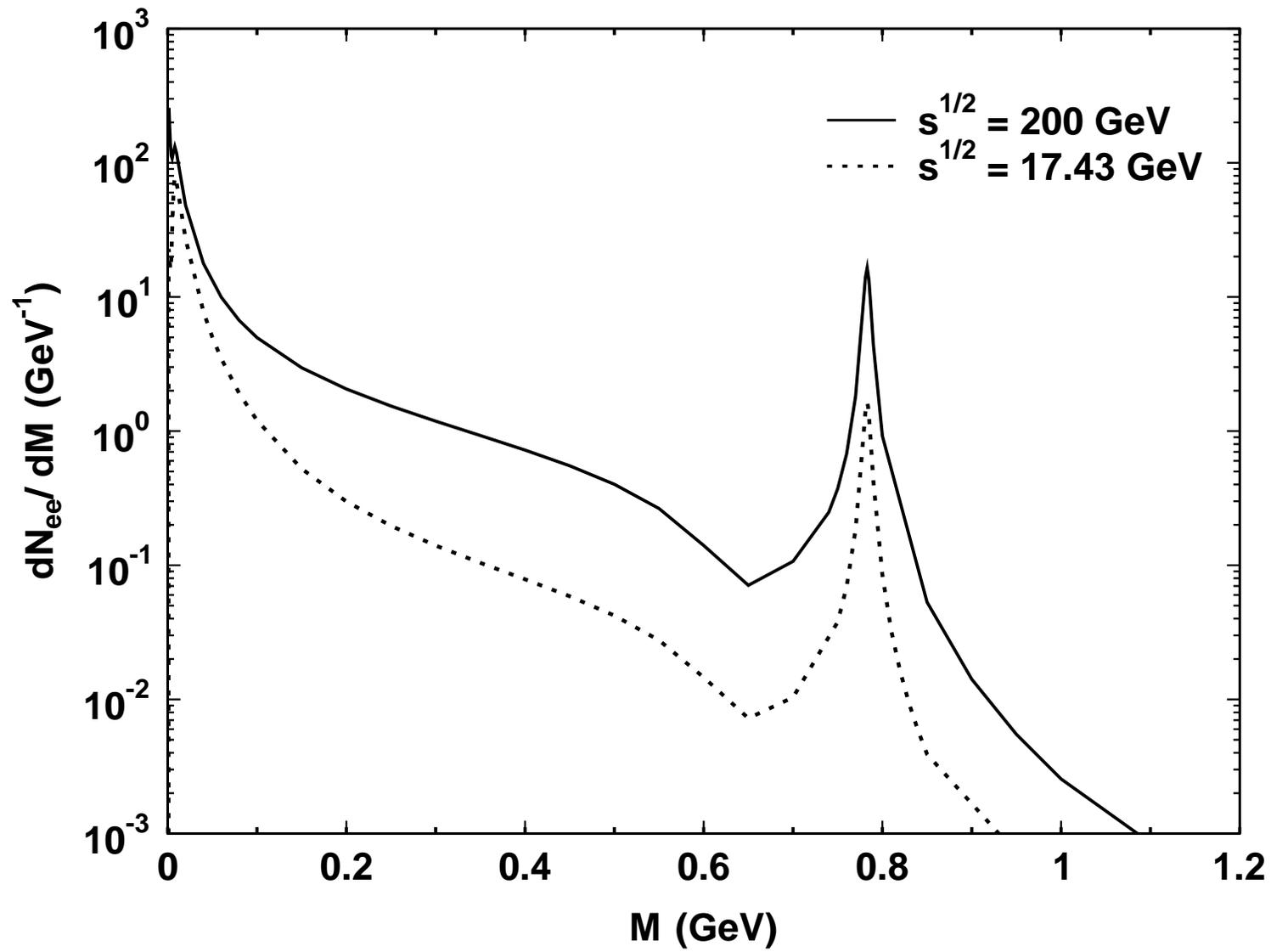
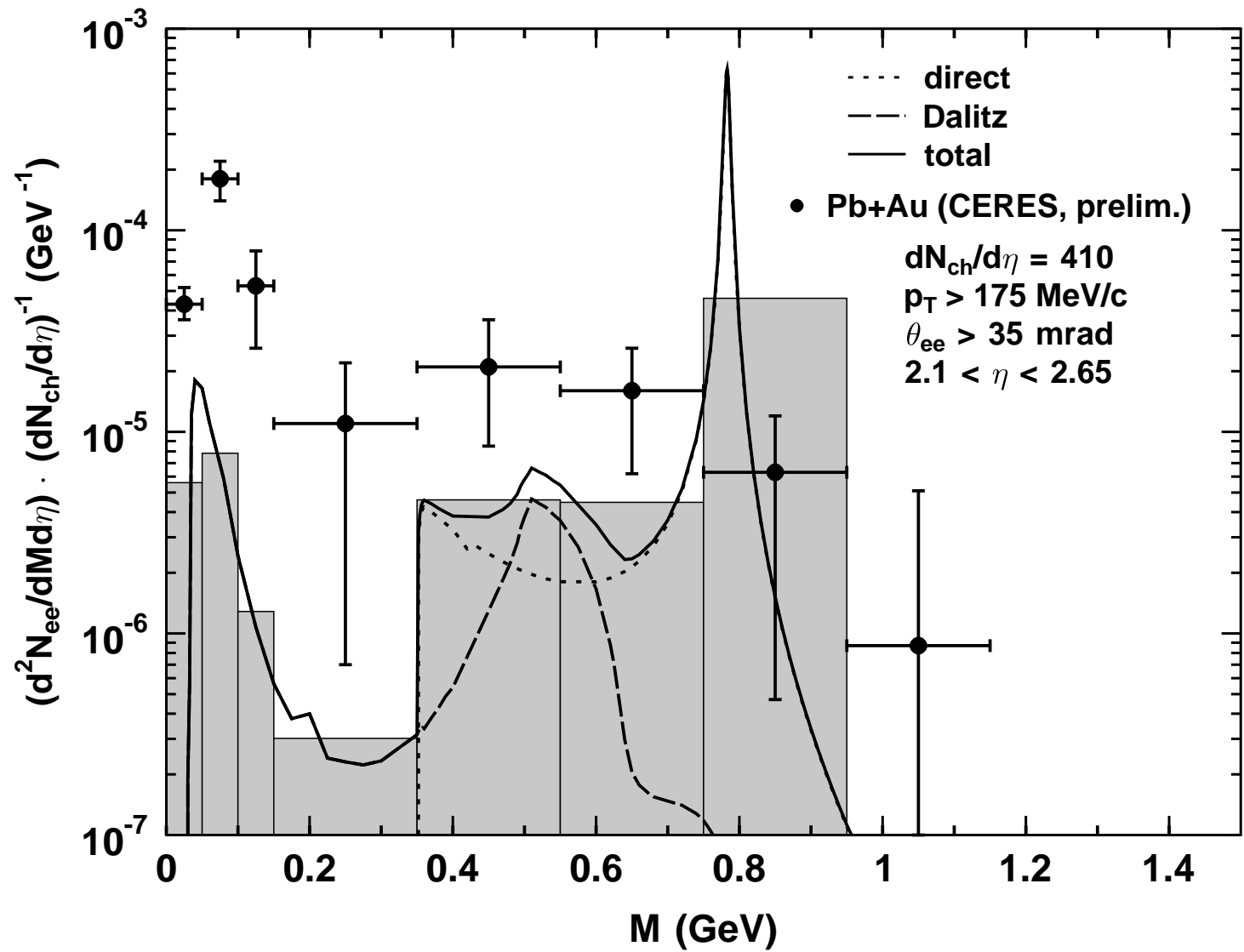
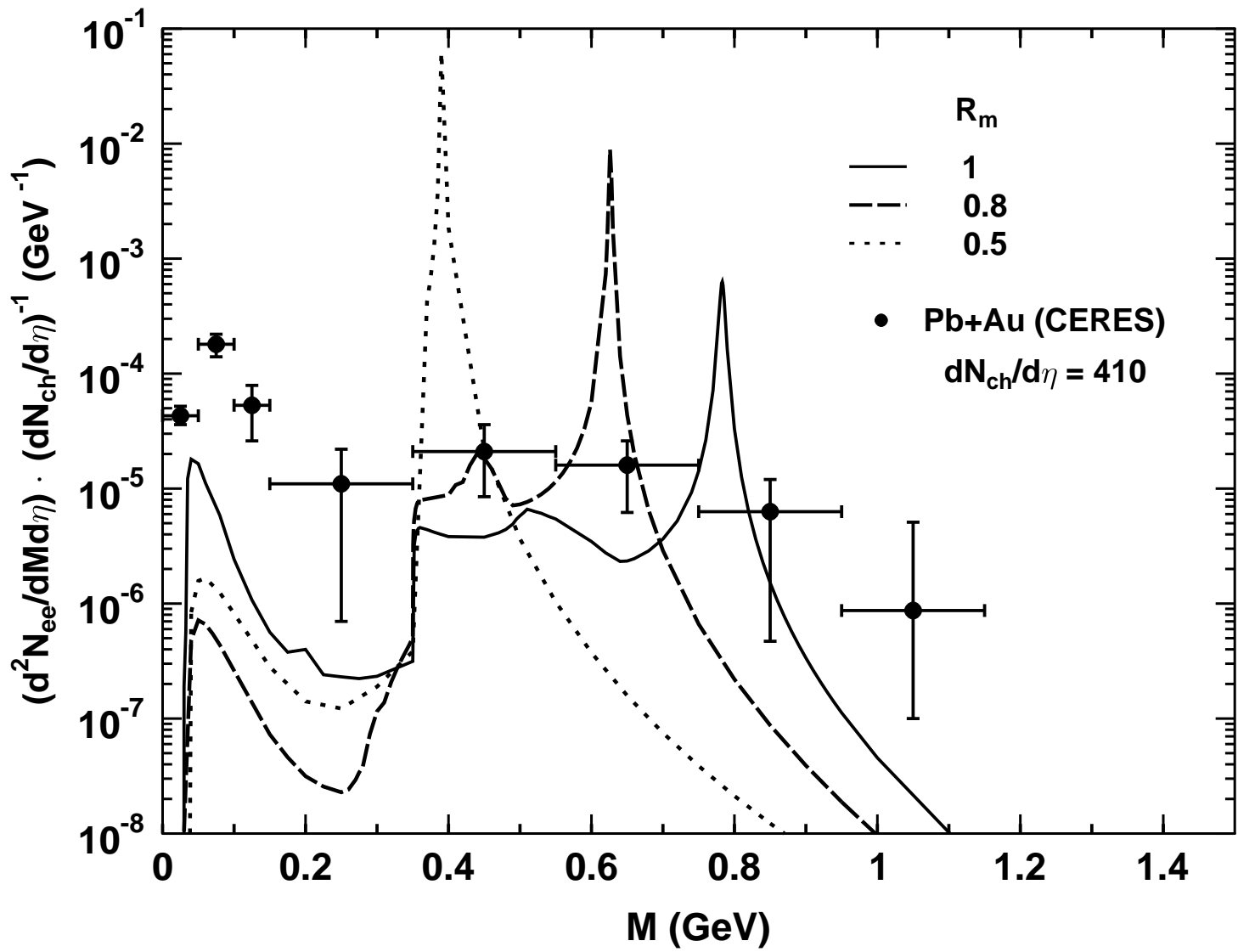
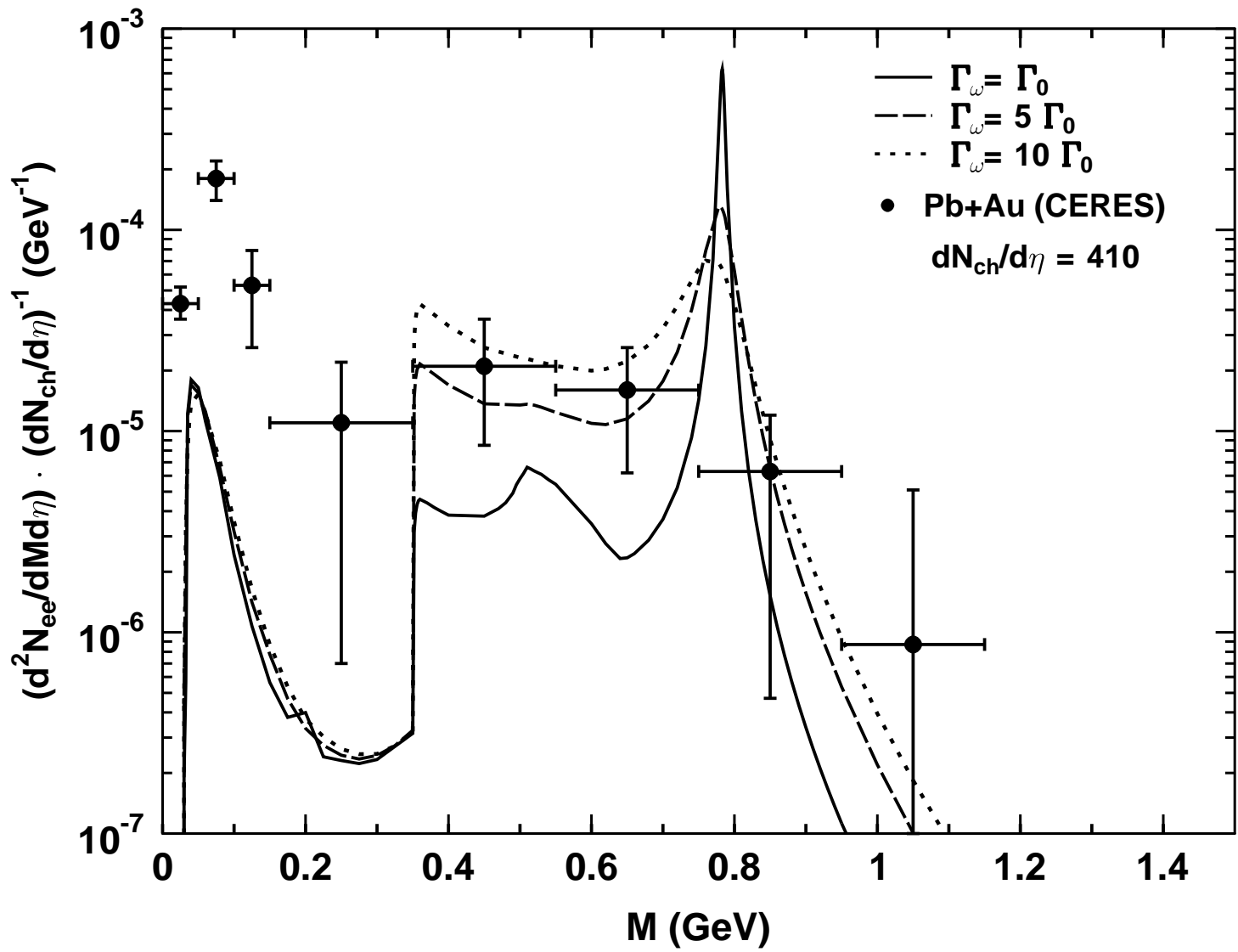
\end{document}